\documentclass{ws-procs9x6}

\begin{document}

\title{DOPING: A NEW NON-PARAMETRIC DEPROJECTION SCHEME}

\author{DALIA CHAKRABARTY$^*$ and LAURA FERRARESE}

\address{School of Physics $\&$ Astronomy, University of Nottingham,\\
Nottingham, NG7 2RD, U.K.\\
$^*$E-mail: dalia.chakrabarty@nottingham.ac.uk\\
www.university\_name.edu}

\begin{abstract}
We present a new non-parametric deprojection algorithm DOPING
(Deprojection of Observed Photometry using and INverse Gambit), that
is designed to extract the three dimensional luminosity density
distribution $\rho$, from the observed surface brightness profile of
an astrophysical system such as a galaxy or a galaxy cluster, in a
generalised geometry, while taking into account changes in the
intrinsic shape of the system. The observable is the 2-D surface
brightness distribution of the system. While the deprojection schemes
presented hitherto have always worked within the limits of an assumed
intrinsic geometry, in DOPING, geometry and inclination can be
provided as inputs. The $\rho$ that is most likely to project to the
observed brightness data is sought; the maximisation of the likelihood
is performed with the Metropolis algorithm. Unless the likelihood
function is maximised, $\rho$ is tweaked in shape and amplitude, while
maintaining positivity, but otherwise the luminosity distribution is
allowed to be completely free-form. Tests and applications of the
algorithm are discussed.
\end{abstract}


\bodymatter

\section{Introduction}
\label{sec:intro}
\noindent
The preliminary step involved in the dynamical modelling of galaxies,
concerns the deprojection of the observed surface brightness
distribution into the intrinsic luminosity density, as has been
practised by [\refcite{krajnovic04}], [\refcite{kronawitter00}],
[\refcite{magog98}], among others. Deprojection though, is a non-unique
problem, unless performed under very specific configurations of
geometry and inclination, as discussed by [\refcite{binneygerhard}],
[\refcite{kochanekrybicki}], [\refcite{rybicki87}],
[\refcite{vandenbosch}] and others.

Over the years, several deprojection schemes have been advanced and
implemented within the purview of astronomy; these include parametric
formalisms designed by [\refcite{bendinelli}], [\refcite{palmer}] and
[\refcite{cappellari}], as well as non-parametric methods, such as the
Richardson-Lucy Inversion scheme developed by [\refcite{richardson}] and
[\refcite{lucy}] and a method suggested by [\refcite{romkoch}]. While the
parametric schemes are essentially unsatisfactory owing to the
dependence of the answer on the form of the parametrisation involved,
the non-parametric schemes advanced till now have suffered from the
lack of transparency and in the case of the Richardson-Lucy scheme,
lack of an objective convergence criterion.

Here, we present a new, robust non-parametric algorithm: Deprojection
of Observed Photometry using an INverse Gambit (DOPING). DOPING does
not need to assume axisymmetry but can work in a triaxial geometry
with assumed axial ratios, and is able to incorporate radial
variations in eccentricity. Although the code can account for changes
in position angle, this facet has not been included in the version of
the algorithm discussed here. Also, here we present the 1-D results
obtained with DOPING but the code provides the full 3-D density
distribution that projects to observed brightness map. In a future
contribution, (Chakrabarty $\&$ Ferrarese, {\it in preparation}),
DOPING will be applied to recover the intrinsic luminosity density of
about 100 early type galaxies observed as part of the ACS Virgo
Cluster Survey, as reported in [\refcite{coteacs04}].

The paper has been arranged as follows. The basic framework of DOPING
is introduced in Section~\ref{sec:method}. This is followed by a short
discourse on a test of the algorithm. An application to the observed
data of the galaxy vcc1422 is touched upon in
Section~\ref{sec:1422}. Another application of DOPING is discussed in
Section~\ref{sec:clusters}. The paper is rounded up with a summary of
the results.
\section{Method}
\label{sec:method}
\noindent
The outline of the methodology of DOPING is presented below.
\begin{enumerate}
\item The plane of the sky ($x-y$ plane) projection of the galaxy is
considered to be built of the observed isophotes that we consider to
be concentric and {\it analytically expressible in terms of $x$ and
$y$}, such that the $i^{th}$ isophote has an extent of $a_i$ along
${\bf{\hat{x}}}$ (say). The inputs to DOPING are the brightness $I_i$
and the projected eccentricity $e_p^i$ that define the $i^{th}$
isophote, where $i\in{N}$, $i\leq{N_{data}}$.
\item We set up $\rho=\rho[\xi(x,y,z)]$, where $\xi$ is the
ellipsoidal radius for the geometry and inclination of choice.
\item We identify pairs of $(x_i, y_i)$ that sit inside the elliptical
annulus between the $i^{th}$ and the $i+1^{th}$ isophotes. 
\item At the beginning of every iterative step, the density
distribution $\rho(x_i, y_i, z)$, over the line-of-sight coordinate
$z$, is updated in size and amplitude, $\forall{i}$, subject to the
only constraints of positivity. The scales over which this updating is
performed are referred to as $scl_1$ and $scl_2$.
\item This updating is continued till the maxima in the likelihood is
identified by the inbuilt Metropolis algorithm; the likelihood is
maximised when the observed brightness distribution is closest to the
projection of the current choice of the density. Regularisation is
provided in the form of a penalty function that is set as the product
of the smoothing parameter $\alpha$ and a function of the Laplacian of
$\rho$.  
\item The spread in the models in the neighbourhood of the maximal
region of the likelihood function, is used to formulate the
($\pm$1-$\sigma$) errors on the estimated density.
\end{enumerate}
\section{Tests}
\label{sec:test}
\noindent
Prior to the implementation of the algorithm, it is extensively tested
using analytical models. For one of these test, the results of which
are presented in Figure~1, the surface brightness distribution is
extracted by projecting the analytically chosen luminosity density
distribution of a toy oblate galaxy. The projection is performed along
the LOS coordinate $z$, under the assumption of an intrinsic minor to
major axis ratio that goes as $\sqrt{1/(1+r^2/r_c^2)}$, (where $r$ is
the spherical radius and the scale length $r_c$ is 0$^{''}$.5). Thus,
by construction, this toy galaxy is rounder inside the inner
0$^{''}$.5 and outside this radius, it quickly (by 3$^{''}$) flattens
to a disky system with eccentricity of about 0.99.

This toy galaxy is viewed edge-on (at 90$^\circ$) for this test. The
brightness distribution is then ported to DOPING and the deprojection
is carried out under a chosen geometry+inclination configuration; the
recovered luminosity density is compared to the true density of this
model.

In Figure~1, the true density of this toy oblate galaxy is shown in
open circles, along the photometric major (left panel) and minor
(right) axes. The open triangles are used to depict the density
recovered by DOPING, under the assumptions of face-on viewing angle
($i$) and triaxiality, with the LOS extent set to double the
photometric major axis. Similarly, when the galaxy is viewed at
$i$=20$^\circ$, with the LOS extent set to half the photometric major
axis, the recovered density distribution is plotted along the azimuths
of 0$^\circ$ and 90$^\circ$ in crosses. In both cases, the two
photometric axes are set as related in the way suggested by the
projected eccentricity ($e_p$) data. In the former case, the value of
$\alpha$ that is used in the penalty function, is 10 times higher than
in the latter case.

As expected, when the LOS extent is set longer than in the test
galaxy, it leads to a smaller density than the true density while a
shorter LOS extent is betrayed in higher recovered densities. Also,
when the test galaxy, modelled as triaxial, is not viewed along one of
the principle axes, then as expected, isophotal twist is
recovered. This results in a steeper drop in the projection of the
recovered density in the case of $i$=20$^\circ$ (not shown here) than
in the brightness data.

Like all other deprojection algorithms, DOPING also requires a seed or
a trial density distribution to begin with. The parameters in the
Metropolis algorithm, namely the temperature and the scale lengths
$scl_1$ and $scl_2$ (see Section~\ref{sec:method}), are chosen to ensure
robustness of the algorithm.

\begin{figure}
\centering{
\psfig{file=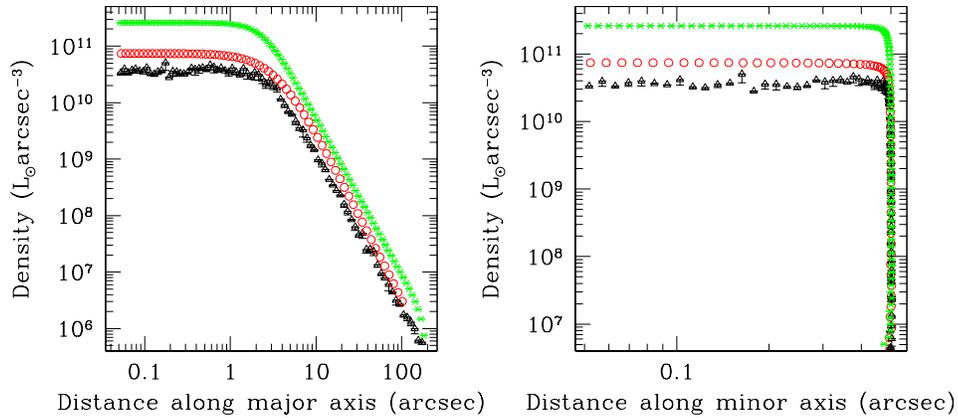,width=5in}
\vskip-6.5cm
\caption{Deprojection of a model brightness profile under assumptions
of (i) ratio of extent along LOS to major axis coordinate=2;
$i$=0$^\circ$ (ii) ratio of LOS extent to extent along ${\bf{\hat x}}$
= 0.5; $i$=20$^\circ$. The recovered density distributions are shown at
azimuths of 0$^\circ$ (right panel) and 90$^\circ$ (left panel), in
crosses for case (ii) and open triangles for case (i). The true
density of this toy galaxy is shown in open circles. The degeneracy in
the deprojection exercise, as a function of inclination and geometry is
brought out in this figure.  }}
\label{fig:test}
\end{figure}

\section{Applications}
\label{sec:1422}
\noindent
We demonstrate the applicability of DOPING in real galaxies by
deprojecting the surface brightness profile of vcc1422 (IC3468), which
is a Virgo Cluster dwarf elliptical, covered by the ACS Virgo Cluster
Survey\cite{laura06}. We choose this galaxy since being about 8
magnitudes fainter at the centre than the test galaxy considered
above, it illustrates the efficacy of DOPING over a wide range of
brightness. Photometrically, it is evident that this galaxy has a
small central component, (a nucleus extending to about 0$^{''}$.3),
that sits on top of a more extended component. The projected
ellipticity of this outer component meanders its way up from about
0.12 at about 0$^{''}$.4 to 0$^{''}$.3 at about 1$^{''}$.6, to jiggle
down to about 0.22 at about 120$^{''}$.

An experiment was conducted to bring out the importance of including
this information about the variation in $e_p$. To ease such an
exercise, the contribution of the nucleus to the surface brightness
measurement was subtracted and the resulting brightness profile was
then deprojected under different conditions. When all this variability
in the projected shape of the galaxy is incorporated into the
deprojection technique, the density recovered along the photometric
major axis ${\bf{\hat x}}$ is depicted in open circles in
Figure~2. This is compared to the density obtained under the
assumption that the whole galaxy admits a single $e_p$ of 0.25. In
both cases, oblateness and edge-on viewing are assumed. Since the two
profiles significantly differ, the comparison brings out the
importance of including the details of the variation in the
eccentricity, even for this mildly eccentric system. With a more
radically varying $e_p$ profile, this difference would only increase!

DOPING has the capacity for deprojecting a multi-component system such
as vcc1422; in these cases, the seed for the sought density is chosen
such that it reflects the existence of all the components. Thus, in
the case of vcc1422, inside 0$^{''}$.3, the seed should bear
signatures of both the components, while outside it, only the
contribution from the more extended of the two components is
required. The result of this deprojection is shown in
Figure~3. The deprojection was performed under
the assumptions of oblateness and an edge-on inclination.

\begin{figure}
\centering{
\psfig{file=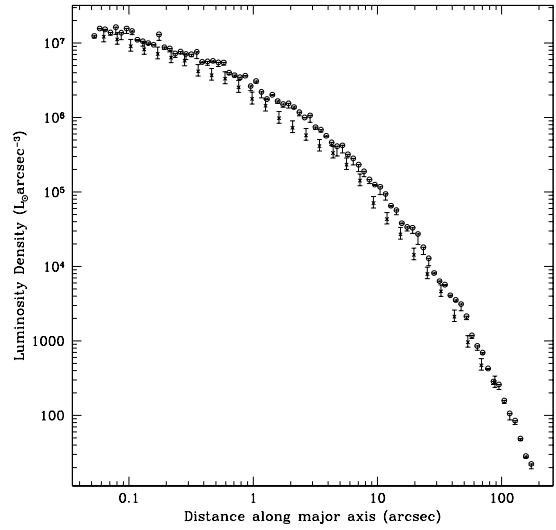,width=3in}
\caption{The recovered density distribution of the nucleus subtracted
brightness distribution of vcc1422, plotted along the ${\bf{\hat x}}$,
obtained under the assumptions of $i$=90$^\circ$ and oblateness. The
profile obtained from including the information about the changes in
$e_p$ with $x$ is marked with open circles, while the deprojection
carried out under the assumption of a constant $e_p$ of 0.25 is
represented in crosses.}}
\label{fig:1422}
\end{figure}

\begin{figure}
\centering{
\psfig{file=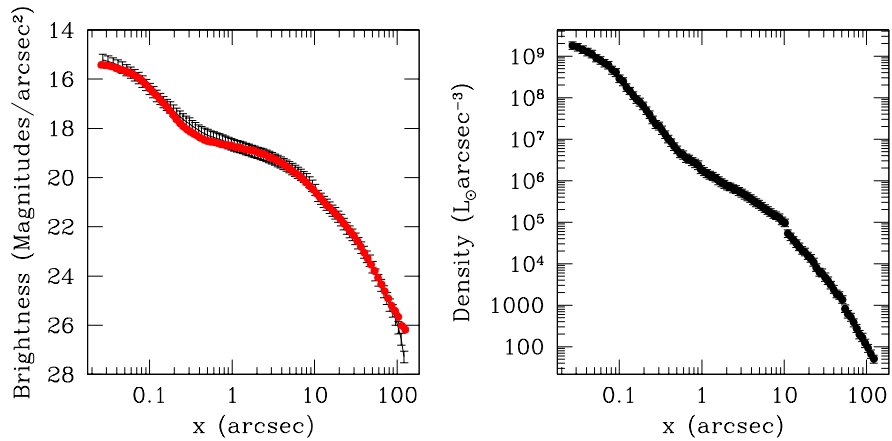,width=5in}
\vskip-6.5cm
\caption{Deprojection of the surface brightness data of the nucleated
galaxy vcc1422. The recovered density is presented in the right while
its projection has been overlaid on the observed brightness profile
(in grey). Again, an oblate geometry and edge-on viewing were
adopted.}}
\label{fig:1422_nucleus}
\end{figure}

\section{3-D Morphology of Galaxy Clusters}
\label{sec:clusters}
\noindent
A project is underway (Chakrabarty, de Philippis $\&$ Russell {\it in
preparation}) to decipher the true intrinsic morphology and
inclination (to the LOS) of a galaxy cluster, by deprojecting its
X-ray brightness distribution under distinct assumptions about the
cluster geometry and orientation; the deprojection in question is
carried out by DOPING, using the measured projected eccentricity of
the system ($e_p$). If available, information about the LOS extent of
the cluster is also implemented. Such information is attainable for
galaxy clusters since the hot gas in these systems is capable of
scattering the primordial CMB photons; this is referred to as the
Sunyaev-Zeldovich Effect (SZe). 

The cluster morphology is identified as oblate, prolate or triaxial
from the mutual weighing of the different deprojected density profiles
that are recovered under assorted deprojection scenarios; the
deprojected density distributions are sought along the
${\bf{\hat{x}}}$ axis. The availability of the SZe measurements
allows for marked tightening of the constraints that are placed on the
inclination of a system from the analysis of the X-ray brightness
information alone. For the triaxial systems, the SZe data can also
help constrain the intrinsic axial ratios.

In Figure~4, the recovered density profiles for the Abell clusters
A1651 and A1413 are depicted; our analysis indicates that A1651 is
prolate while A1413 is triaxial, with intrinsic axial ratios of 0.96
and 1.64. The inclinations $i$ are found to be $8^\circ < i <
32^\circ$ for A1651 and $66^\circ < i < 71^\circ$ for A1413.

\begin{figure}
\centering{
\psfig{file=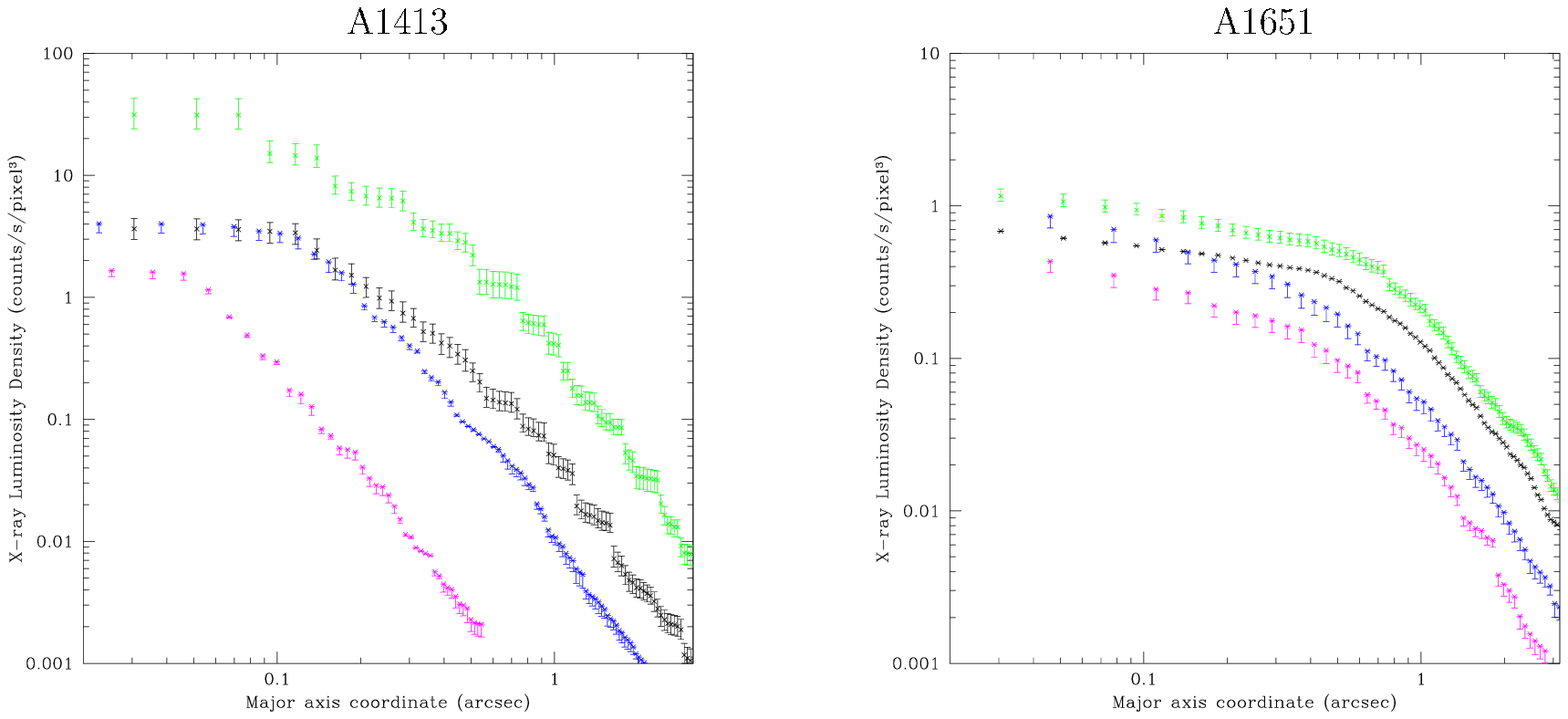,width=5.8in}}
\caption{Density profiles along ${\bf{\hat{x}}}$, recovered by
deprojecting the {\it Chandra} X-ray brightness distribution of
clusters Abell~1651 (right) and Abell~1413 (left), under the
assumptions of (i) oblateness and $i$=90$^\circ$, shown in crosses
(ii) oblateness and $i$=$i_{min}$; in filled circles (iii) prolateness
and $i$=90$^\circ$; in filled triangles (iv) prolateness and
$i$=$i_{min}$; in open circles. Here $i_{min}$ is the smallest
inclination allowed under oblateness, for a measured (uniform) $e_p$
(=$\sin^{-1}e_p$).}
\label{fig:clusters}
\end{figure}

\section{Summary}
\noindent
In this paper, we have introduced a new non-parametric algorithm
DOPING that is capable of inverting observed surface brightness
distributions of galaxies and galaxy clusters, while taking into
account variations in the intrinsic shapes of these systems. The
potency of DOPING is discussed in the context of a test galaxy in
which the eccentricity is made to change radically with radius. The
code is also successfully applied to obtain the luminosity density
distribution of the faint nucleated galaxy vcc1422. Lastly, a novel
use is made of the capability of DOPING to deproject in general
geometries, in determining the intrinsic shape and inclination of a
galaxy cluster. It is envisaged that implementing a measure of the LOS
extent of a cluster from Sunyaev Zeldovich measurements, will help
tighten the estimates of cluster inclination and the intrinsic axial
ratios of triaxial clusters.

\end{document}